\documentclass[preprint,12pt]{elsarticle}
\usepackage{amssymb}
\usepackage{amsmath}

\usepackage{hyperref}
\usepackage{xurl}
\usepackage{float}
\usepackage{color}

\journal{}

\begin{document}

\begin{frontmatter}

\title{Opinion dynamics model of collaborative learning}

\author[1,2]{Jibeom Seo} 
\author[1]{Beom Jun Kim\corref{cor1}} 
\ead{beomjun@skku.edu}
\cortext[cor1]{Corresponding author}
\address[1]{Department of Physics, Sungkyunkwan University, Suwon 16419, Republic of Korea}
\address[2]{Department of Science Education (Physics Major), Seoul National University, Seoul 08826, Republic of Korea}

\begin{abstract}
We propose a simple model to explore an educational phenomenon where the correct answer emerges from group discussion. We construct our model based on several plausible assumptions: (i) We tend to follow peers' opinions. However, if a peer's opinion is too different from yours, you are not much influenced. In other words, your opinion tends to align with peers' opinions, weighted by the similarity to yours. (ii) Discussion among group members helps the opinion to shift toward the correct answer even when the group members do not know it clearly. However, if everyone tells exactly the same, you often get lost and it becomes more difficult to find the correct answer. In other words, you can find the correct answer when everyone has largely different voices. (iii) We are sometimes stuck to our past. If you keep one opinion for a long time, such a memory works like an inertia in classical mechanics. We use our model to perform numerical investigations and find that the performance of a group is enhanced when initial opinions are diverse, that a lower memory capacity makes consensus occur faster, and that a small group size, typically three or four, is beneficial for better group performance. 
\end{abstract}


\end{frontmatter}

\section{Introduction}
Student learning is a complex dynamic process influenced by a variety of factors including teachers, classmates, textbooks, and even geographical context such as the country of residence~\cite{organisation2010pisa}. Extensive research has been conducted in various fields, such as educational psychology and educational technology, to reveal factors for more effective learning~\cite{rutten2012learning, weinstein2018teaching}. Additionally, physicists have introduced mathematical models based on educational psychology theories to analyze student data~\cite{pritchard2008mathematical, nitta2010mathematical}. These physics-inspired models use master equations to predict how test scores improve from pretest scores. Notably, models following constructivist theories, which claim that students learn by integrating experiences and social interactions with their existing knowledge, have explanatory power. However, since these studies are based on the mean-field approach, their models have limitations in capturing individual and group characteristics. 

Agent-based modeling (ABM) is another method used to study social learning processes within classrooms. Combined with computer simulations, it allows for the description of macroscopic social phenomena by modeling the microscopic components of social systems~\cite{castellano2009statistical}.
Physicists have applied ABM to investigate social learning (i.e., students' learning) using physics concepts, such as the generalized Ising model~\cite{bordogna2001theoretical,bordogna2002cellular,bordogna2003simulation} and the kinetic theory of gases~\cite{burini2016collective,ormazabal2021agent}.
These studies assume that students interact with their environment--—including teachers, peers, or bibliographic resources--—based on the models’ rules. For instance, a student's idea may change as a result of cognitive influences arising from interactions with teachers, peers, and other information resources, and these impacts depend on the student's current knowledge and characteristics. Key findings identified across these studies include: (1) Combination of lectures, group work, and individual study is more effective than attending lectures alone. (2) Students tend to perform better in heterogeneous groups with diverse ideas compared to homogeneous groups with similar ideas. (3) Excessively large group sizes may negatively affect student achievements.

However, one study~\cite{lee2022effect} on peer-collaborated activities in a physics course reveals interesting results that contradict our common sense: Groups in which none of the members initially know the correct answer can reach the correct answer after discussion, but only if their initial answers are not identical. These results were achieved solely through peer work without teacher assistance, bibliographies, etc.
Unlike the ABM studies mentioned above, where teacher assistance and additional materials are essential to explain improved performance, the new findings indicate that students can even develop without these factors.
Moreover, while previous ABM research has shown slight performance improvements in heterogeneous groups compared to homogeneous ones~\cite{bordogna2001theoretical, burini2016collective}, this study presents substantial differences: Groups with entirely different incorrect answers have a $38.78\%$ chance of reaching the correct answer; groups, where some incorrect answers are different, have a $27.47\%$ chance; and groups, where all incorrect answers are the same, have a $0\%$ success rate. Furthermore, numerous studies on collaborative learning~\cite{singh2005impact,smith2009peer,brundage2023peer} show similar results, supporting that the above-mentioned phenomenon is not exceptional.
These discrepancies between previous ABM studies and empirical educational studies suggest that existing models cannot fully account for the empirical findings, thus necessitating the proposal of a novel modeling approach.

In this study, we aim to model the phenomenon of correct answer emergence using our new agent-based model and to explore the explainability of other educational and psychological principles within the model framework.
In our model, we emphasize the importance of student-student interaction while excluding external factors such as teacher assistance and informational resources. This decision is driven by observations~\cite{lee2022effect} that improvements in student performance during peer-collaborated activities were achieved solely through peer interactions.
Although this decision may seem ambitious, it is supported by Vygotsky’s theory~\cite{vygotsky1978mind}, which asserts that students learn through social interaction. In collaborative group discussions, peers provide each other with the zone of proximal development, thereby expanding the range of tasks a learner can accomplish with peer support (i.e., scaffolding). In other words, even without considering external factors such as teacher assistance, we can speculate that peer support plays a role similar to teacher guidance.
Moreover, earlier ABM studies assume that students are aware of both their own and others' knowledge levels (scored from $1$ for the highest to $-1$ for the lowest) and that those with a higher level of knowledge can effectively elevate their peers' understanding.  
However, these assumptions raise two major issues: First, during peer discussions, it is difficult for students to accurately assess their own or others’ knowledge levels; instead, they can only discern the similarity of opinions. Second, even low-achieving students can positively influence the learning of their peers, and even students' incorrect answers can be useful in problem-solving~\cite{lee2022effect,smith2009peer}. Therefore, in our study, we assume that students are unaware of their own knowledge levels during discussions and can only perceive opinion differences among their peers—--differences that may enhance their performance. For details, see Section~\ref{sec:Model}.

Finally, in Section~\ref{sec:Results}, we will demonstrate how our model successfully simulates the phenomenon where the correct answer emerges from incorrect initial opinions. Additionally, we will discuss how group sizes may affect the group performance, and analyze that initial opinion diversity is more critical than initial group performance in determining final group performance.

\section{Model}\label{sec:Model}
\label{model}
In this section, we propose a model that describes the pedagogical phenomenon of reaching the correct answer from incorrect initial opinions through a series of group discussions. Group members collaboratively discuss the problem and adjust their opinions influenced by discussion. Based on our review of previous ABM studies, educational observations, and pedagogical theories, we propose a model with several assumptions for student-student interaction: (1) Students' opinions are more influenced by others’ opinions that resemble theirs. (2) Diversity of opinions helps the group to find the correct answer. (3) Students' memories of specific topics influence the possibility to revise their opinions. (4) Students initially tackle a given problem individually and form their own initial opinions. It is important that the problem must be factual and concrete and thus answers to it can be scored. (5) Any student within a group is represented only by the opinion they have; no student is considered to be special and thus all students impact others equally.

The opinion of the $i$th student at time $t$ is represented as a three-dimensional unit vector $\vec{S_i}(t)$, which implies that all opinions have the same magnitude, consistent with the above assumption (5). We also define that the unit vector corresponding to the correct answer in three-dimensional space is represented as ${\hat z}=(0,0,1)$, which allows us to assign the score or the performance of the opinion $\vec{S_i}(t)$ as ${\hat z}\cdot \vec{S_i}(t) = S_i^z(t)$. In other words, the performance of a student is measured by the cosine similarity between two vectors, i.e., the opinion of the student and the correct answer. We note that different opinions can share the same score, as long as the $z$-components of the opinion vectors are the same.

\begin{figure}[t]
    \centering
    \includegraphics[width=0.9\textwidth]{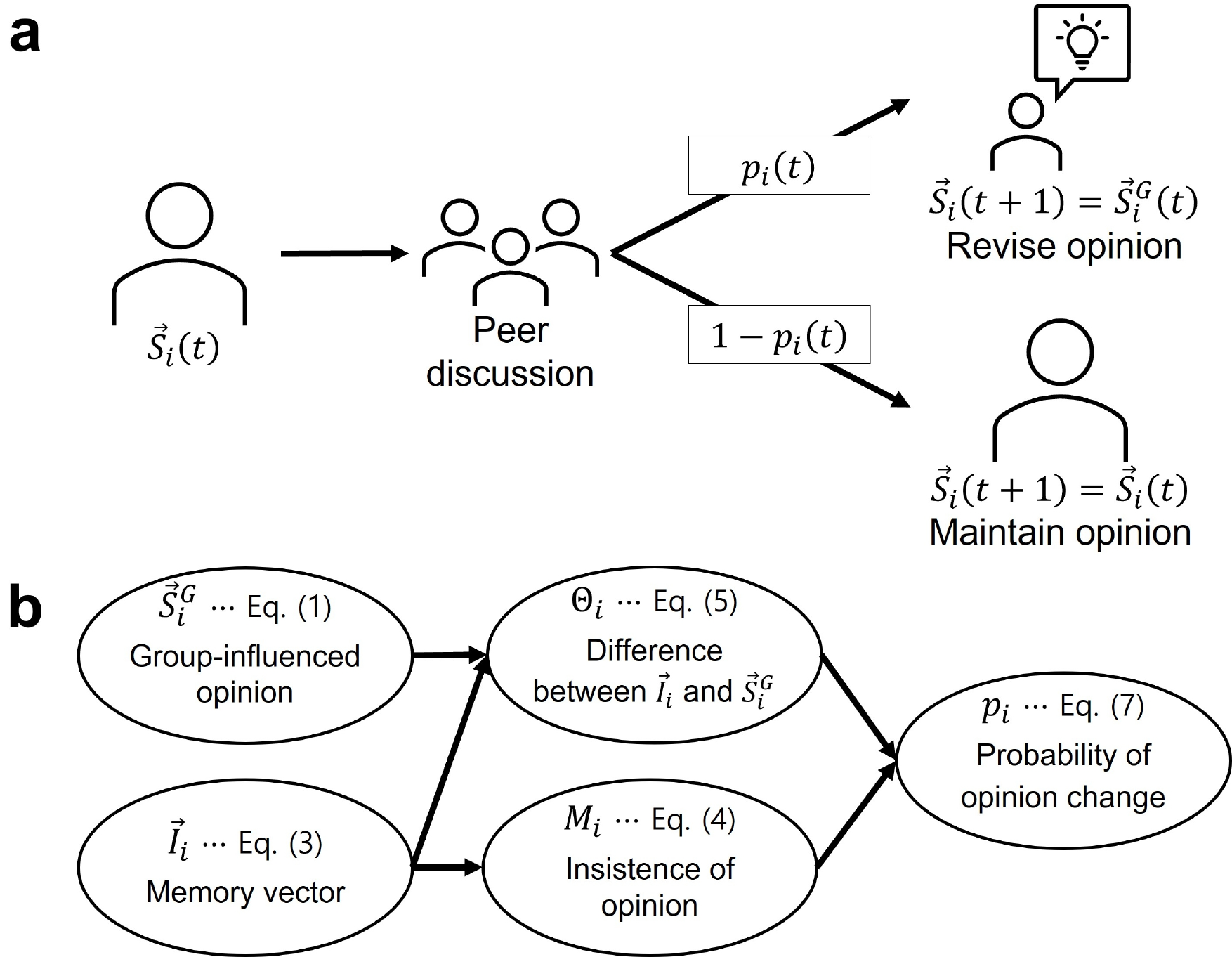}
    \caption{
        (\textbf{a}) Sketch of the model and (\textbf{b}) diagram illustrating the components of opinion change probability in a directed acyclic graph. After initializing all students' opinions, they discuss with their peers at each time step. Students either revise or maintain their opinions based on their own previous opinions and group members' opinions. See Sections~\ref{sec:2.1} and \ref{sec:2.2} for details on the components, and refer to Section~\ref{sec:2.3} for the time evolution of opinions.
    }
    \label{fig:DAG}
\end{figure}

We now explain how our model works. At time $t=0$, each student's opinion is initialized in several different ways. For the initial condition denoted as NH (Northern Hemisphere), we randomly spread initial opinions on the surface of the half sphere above $xy$-plane so that $S_i^z(0) > 0$ for $\forall i$. Likewise, SH (Southern Hemisphere) is the initial condition with $S_i^z(0) <0$, and EH (Eastern Hemisphere) with $S_i^x(0) > 0$, for $\forall i$, respectively. Among the three initial conditions we use in this work, the average score at the initial stage ($t=0$) is the highest for NH, the lowest for SH, and EH in the middle with the average score close to null.
After initialization, each student's opinion evolves over time. In this process, several quantities are involved, which we will introduce and explain one by one below. Once all the relevant quantities are introduced, we will describe how they are incorporated into modeling the dynamics of peer discussion. Before diving into the details, we provide both a schematic representation of the model [Fig.~\ref{fig:DAG}(a)] and a diagram illustrating probability of opinion change [Fig.~\ref{fig:DAG}(b)] for better understanding.

\subsection{Group-influenced opinion}\label{sec:2.1}
We first introduce the group-influenced opinion $\vec{S_i^G}(t)$ for the $i$th student in the group of the size $N$ defined as 
\begin{eqnarray}
\vec{S_i^G}(t) & \equiv &\frac{\vec{G_i}(t)}{| \vec{G_i}(t) |} ,\quad \vec{G_i}(t) \equiv \sum_{j=1}^N \left( \frac{1+\vec{S_i}(t)\cdot\vec{S_j}(t)}{2} \right) \vec{S_j}(t) + {\bar V}N^\alpha \hat{z}, \label{eq:SG} \\
{\bar V} & \equiv & \frac{V}{V_{\rm max}} ,\quad V(\{\vec{S}(t)\}) \equiv \sum_{j \geq i}|\vec{S_i}(t)-\vec{S_j}(t)|^\beta. \label{eq:V} 
\end{eqnarray}
Note that $\vec{S_i^G}(t)$ is also a unit vector like the opinion vector of each student. 

The first term of the right-hand side (RHS) of $\vec{G_i}(t)$ in Eq.~(\ref{eq:SG}) describes the total influence of all $j$th students on the $i$th student. From our common experience that we are more influenced by the other's opinion if it is not much different from ours, we assume that $j$ affects $i$ more strongly when the opinion similarity between the two is larger. If $\vec{S_j}$ points a completely opposite direction to $\vec{S_i}$ on the unit sphere, we have $\vec{S_i}\cdot\vec{S_j} = -1$ and $j$ does not affect $i$'s opinion at all. If the similarity is close to the maximum possible value of $\vec{S_i}\cdot\vec{S_j} = 1$, $i$ is heavily influenced by $j$'s opinion. The influence strength $(1+\vec{S_i}\cdot\vec{S_j})/2 \in [0,1]$ thus plays the role of the weight that $i$ takes into account when the student $i$ adjusts their opinion in line with $j$'s opinion. Consequently, the first term of the RHS of $\vec{G_i}(t)$ is a weighted average of ${\vec S_j}$, taking into account the similarity with $i$'s opinion. 

The second term on the RHS of $\vec{G_i}(t)$ in Eq.~(\ref{eq:SG}) represents the scaffolding effect of the group on the $i$th student, reflecting how the zone of proximal development provided by other students helps guide a student toward the correct answer. In other words, even when all group members do not know the correct answer, a group discussion tends to guide them toward it. However, if everyone's opinion is exactly the same, group discussion cannot help them to escape from it and thus they can be stuck to their own wrong identical opinions. 
Furthermore, since $\bar{V}$ is a function of $\vec{S}$, the way opinion vectors $\vec{S}$ interact influences the value of $\bar{V}$, which in turn affects $\vec{S}$. This indicates that mutual influence (the first term) contributes to performance enhancement.
When viewed through the lens of physics, the scaffolding effect guiding group members toward the correct answer ${\hat z}$ in our model plays a role analogous to the external magnetic field in the ferromagnetic Ising model. However, the field strength is an increasing function of the opinion diversity measured by $V(\{\vec{S}(t)\})$: Group discussion can work better when everyone has a different voice. In the context of opinion dynamics models, a mechanism similar to an external field is often introduced (e.g., mass media in~\cite{gonzalez2005nonequilibrium, bordogna2007dynamic}). Such an external source is typically distinguished from interactions between agents and is located outside the internal system where agents live. In contrast, in our model, the second term on the RHS of $\vec{G_i}(t)$ in Eq.~(\ref{eq:SG}) is constructed from students' opinions within a group (the internal system). This term represents collective peer support that emerges from a co-constructed zone of proximal development and functions as a form of teacher guidance. Therefore, we regard it as an external influence, albeit one that originates from within the internal system.

Our definition of the opinion diversity measure $V$ in Eq.~(\ref{eq:V}) is inspired by the Thomson problem~\cite{thomson1904xxiv}, which seeks the minimum energy configuration of $N$ identical electric charges on the surface of a unit sphere. We adapt this physics approach to define the diversity $V(\{\vec{S}(t)\})$ in Eq.~(\ref{eq:V}) as an increasing function ($\beta > 0$) of the opinion difference $|\vec{S_i}-\vec{S_j}|$. In the physicist's view, the parameter $\beta$ determines how the potential between charges is proportional to distance. A suitable normalization of the opinion diversity is also made considering the maximum possible value $V_{\rm max}$ of $V$ for a given number of students in a group. 

The last component in Eq.~(\ref{eq:SG}) that we are to explain is $N^\alpha$. Our choice of $\alpha<1$ signifies a diminishing incremental scaffolding effect as the group size becomes larger. This concept originates from the second principle of Latané's theory~\cite{latane1981psychology}, which suggests that the social impact of multiple sources on an individual is a sublinearly increasing function of the group size. From a physicist's point of view, the parameter $\alpha$ determines how proportional the size of the external field is to the system size $N$.

In summary, the group-influenced opinion vector in our model is formed in two different but related ways: Weighted influence from other students' individual opinions and diversity-related inference toward the correct answer. It is to be noted that the former effect in $\vec{G_i}(t)$ of Eq.~(\ref{eq:SG}) is $O(N)$ while the latter $O(N^\alpha)$ with $\alpha < 1$, which implies that interaction term becomes more significant for larger group sizes. 

\subsection{Memory horizon}\label{sec:2.2}
Following constructivist views and our intuitions, it is plausible that our reasoning depends on our prior knowledge in a broad context. Your most recent ideas affect your current reasoning more, while older thoughts may be replaced or forgotten and thus hardly affect you. To mimic such a memory effect, we define the memory vector of the $i$th student $\vec{I_i}(t)$ as
\begin{eqnarray}
\vec{I_i}(t) & \equiv & \sum^t_{t'=\mbox{max}\left(0,\ t-T+1\right)}\vec{S_i}(t'), \label{eq:I}
\end{eqnarray}
where $T$ plays the role of the memory capacity or memory horizon. We also define the insistence or inertia of opinion $M_i$ $(\in [0,T])$ as
\begin{eqnarray}
M_i (t) & \equiv & \left|\vec{I_i}(t)\right|. \label{eq:M}
\end{eqnarray}
For example, if a student consistently maintains the same opinion over the period $T$, the insistence of opinion $M_i$ becomes the largest, indicating a strong adherence to the initial opinion. We next quantify the difference between the individual memory ${\vec I_i}(t)$ in Eq.~(\ref{eq:I}) and the group opinion $\vec{S_i^G}(t)$ in Eq.~(\ref{eq:SG}) by the angle between the two: 
\begin{equation}
\Theta_i(t)  \equiv  \mbox{arccos}\left(\vec{S_i^G}(t)\cdot\frac{\vec{I_i}(t)}{\left|\vec{I_i}(t)\right|}\right) \in \left[ 0,\pi \right] .
\label{eq:Theta}
\end{equation}

\subsection{Time evolution of opinion}\label{sec:2.3}
Now we explain how the opinion of each student evolves over time. Starting from initial opinion distributions NH, SH, and EH (see above) at $t=0$, the model runs until the final time step $t=t_f$ is reached. The time unit $\Delta t = 1$ in the present work corresponds to the macro time step of Monte Carlo simulation in statistical physics, thereby involving updates of all opinion variables within a group: At each time step, all group members interact with each and everyone in the group. Each student's opinion changes to the group-influenced opinion $\vec{S_i^G}(t)$ in Eq.~(\ref{eq:SG}) with the probability $p_i(t)$; otherwise, with the probability $1-p_i(t)$, it remains unchanged [see Fig.~\ref{fig:DAG}(a)]:
\begin{equation}
\label{eq:evolve}
\vec{S_i}(t+1)=
\begin{cases}
    \vec{S_i^G}(t)& \mbox{with probability }p_i(t) , \\
    \vec{S_i}(t) & \mbox{with probability }1-p_i(t) . 
\end{cases}
\end{equation}
The probability $p_i(t)$ changes based on opinion evolution, influenced by $\vec{S_i^G}(t)$ and $\vec{I_i}(t)$. The stochastic change of the individual opinion incorporates two principles. First, if the opinion has remained unchanged over time, i.e., if $M_i$ in Eq.~(\ref{eq:M}) is large, the probability of following the group-influenced opinion becomes small, reflecting the inertia or insistence effect. Therefore, $p_i$ needs to be a decreasing function of $M_i$. Second, if the difference between $\vec{I_i}(t)$ and $\vec{S_i^G}(t)$ is small, a student is more likely to align with the group's opinion. Therefore, $p_i$ needs to be a decreasing function of $\Theta_i$ in Eq.~(\ref{eq:Theta}). In total, we thus write the probability $p_i$ as follows:
\begin{equation}
p_i(t)=\frac{1}{1+M_i(t)\Theta_i(t)}
\label{eq:pi}
\end{equation}

as a decreasing function of both $M_i$ and $\Theta_i$, with the unity in the denominator included to avoid the divergence at $M_i\Theta_i = 0$. In previous studies adopting bounded confidence models~\cite{deffuant2000mixing,rainer2002opinion}, an agent's opinion changes continuously within a set threshold based on their previous opinion. However, in practice, we often encounter a situation where opinion shifts abruptly in the classroom~\cite{hashweh1986toward,chinn1993role}. If a student's preconception conflicts with both a new scientific conception and a real-world experience, the student could radically change their opinion, despite initial contradictions with their beliefs. Our model allows such an abrupt change as can be seen in Eq.~(\ref{eq:evolve}).  

Lastly, with respect to the temporal dynamics of opinion, we demonstrate how iterative peer discussions contribute to group performance improvement. To analyze this, we focus on the first and second terms in the RHS of $\vec{G_i}(t)$ in Eq.~(\ref{eq:SG}). The first term implies that students do not simply adopt the opinions of their group members. Instead, the influence of each opinion varies depending on the differences between them, which prolongs the time required for opinion convergence. As a result, the duration of opinion disagreement—--or the persistence of opinion diversity within the group—--is extended. This, in turn, suggests that the second term can influence the first term for a longer period. To summarize, the two terms interact in the process of converging to the correct answer, which shows the contribution of mutual influence (the first term) to group performance. Furthermore, although not shown here in detail, we found that when the first term is instead formulated as $\sum_j \vec{S}_j$ (representing that students simply adopt the average opinions of their group members), the final performance decreases and opinion consensus occurs more quickly compared to our original model.

\section{Results}\label{sec:Results}
\subsection{Temporal behavior of group performance and opinion diversity}\label{sec:3.1}

\begin{figure}[t]
    \centering
    \includegraphics[width=1\textwidth]{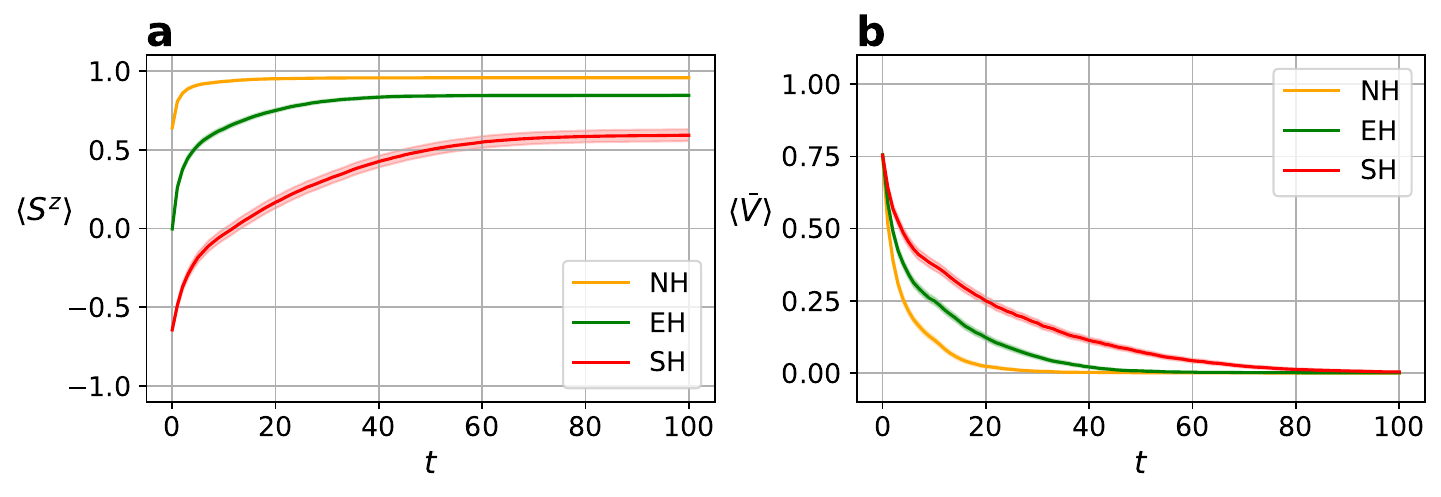}
    \caption{
        Time evolution of (\textbf{a}) group performance $\langle S^z(t) \rangle$ and (\textbf{b}) the opinion diversity $\langle {\bar V}(t)\rangle $ for initial conditions NH (Northern Hemisphere, high score), EH (Eastern Hemisphere, middle score), and SH (Southern Hemisphere, low score) with group size $N=4$, $\alpha=\beta = 0.5$, $T=10$, and $t_f=100$. Here, $\alpha$ represents the strength of the incremental scaffolding effect as group size increases, while $\beta$ determines the sensitivity to opinion difference in measuring opinion diversity; lower $\beta$ results in amplified sensitivity to small opinion differences. $T$ and $t_f$ denotes memory capacity and the final time step of the model simulation, respectively. Ensemble averages $\langle \cdots \rangle$ are taken over $1,000$ independent runs. Solid lines represent the averages, while the shaded regions denote the confidence intervals corresponding to two times the standard deviation ($2\sigma$). It is shown clearly that for all initial conditions (NH, EH, and SH), the group performance increases and the opinion diversity decreases as group discussions proceed. 
    }
    \label{fig:IC}
\end{figure}

In Fig.~\ref{fig:IC}, we first report the temporal behavior of the average group performance measured by $\langle S^z(t) \rangle \equiv \langle \sum_i S_i^z(t)/N \rangle$ [Fig.~\ref{fig:IC}(a)], and the average opinion diversity $\langle {\bar V}(t) \rangle$ with ${\bar V}(t)$ in Eq.~(\ref{eq:V}) [Fig.~\ref{fig:IC}(b)], for $N=4$, $\alpha = \beta = 0.5$, and $T=10$ in Eqs.~(\ref{eq:SG}), (\ref{eq:V}), and (\ref{eq:I}), respectively. We randomly generate $\vec{S}_i(t=0)$ based on a given initial condition among NH, EH, and SH, and then let the system evolve following Eqs.~(\ref{eq:evolve}) and (\ref{eq:pi}) until the final time $t_f = 100$ is reached. We repeat the whole process to perform the ensemble averages $\langle \cdots \rangle$ over 1,000 independent runs.

For the NH initial condition, where all initial opinion vectors are put on the northern hemisphere ($S_i^z(0) > 0$), the group performance $\langle S^z(t) \rangle$ is found to be the highest, which is followed by EH and SH in the descending order, as shown in Fig.~\ref{fig:IC}(a). Our simulation results imply that the initial group performance is important to reach the correct answer more closely: The larger $\langle S^z(0) \rangle$ is, the better the final group performance $\langle S^z(t_f) \rangle$  is.  It is also important to note that regardless of initial conditions, all three (NH, EH, and SH) clearly exhibit improvement in group performance, which implies that group discussions help the students to get to the correct answer as the discussion sessions are iterated. Figure~\ref{fig:IC}(b) for the opinion diversity also displays interesting and important findings: Opinion diversity unanimously decreases for all initial conditions (NH, EH, and SH), and at the final stage of discussion the diversity vanishes, which indicates that all group members eventually achieve the consensus state.

In our investigations, we set $\alpha=0.5$, which leads to group performance improvement in a manner consistent with empirical observations. Although not shown here, we have verified that the final group performances with both $\alpha \rightarrow 0$ and $\alpha \rightarrow 1$ contradict real-world results: Regardless of initial conditions, the former shows too little improvement in group performance throughout group discussions, while the latter shows that all groups achieve the perfect score after all.

Similarly, throughout the present investigation, the parameter $\beta$ in the definition of the opinion diversity $V$ is set to $\beta=0.5$.
Our choice of $\beta < 1$ makes the opinion diversity $V$ more sensitive to smaller opinion differences but less sensitive to larger ones, consistent with the model's assumption (1). This can be understood intuitively: When $\beta < 1$, the function $|\vec{S_i}-\vec{S_j}|^\beta$ becomes concave. Similarly to this function, any increasing concave function $f(x)$ for $x\in[0,2]$ shows greater sensitivity to small changes in $x$ at small $x$ compared to large $x$.
However, both too small ($\beta \rightarrow 0$) and too large ($\beta > 1$) values of $\beta$ are unrealistic. In the former case, the system becomes overly sensitive to even small opinion differences (analogous to $\alpha \rightarrow 1$), while in the latter case, it becomes too insensitive (analogous to $\alpha \rightarrow 0$).

\subsection{Effect of memory horizon}

\begin{figure}[t]
    \centering
        \includegraphics[width=1\textwidth]{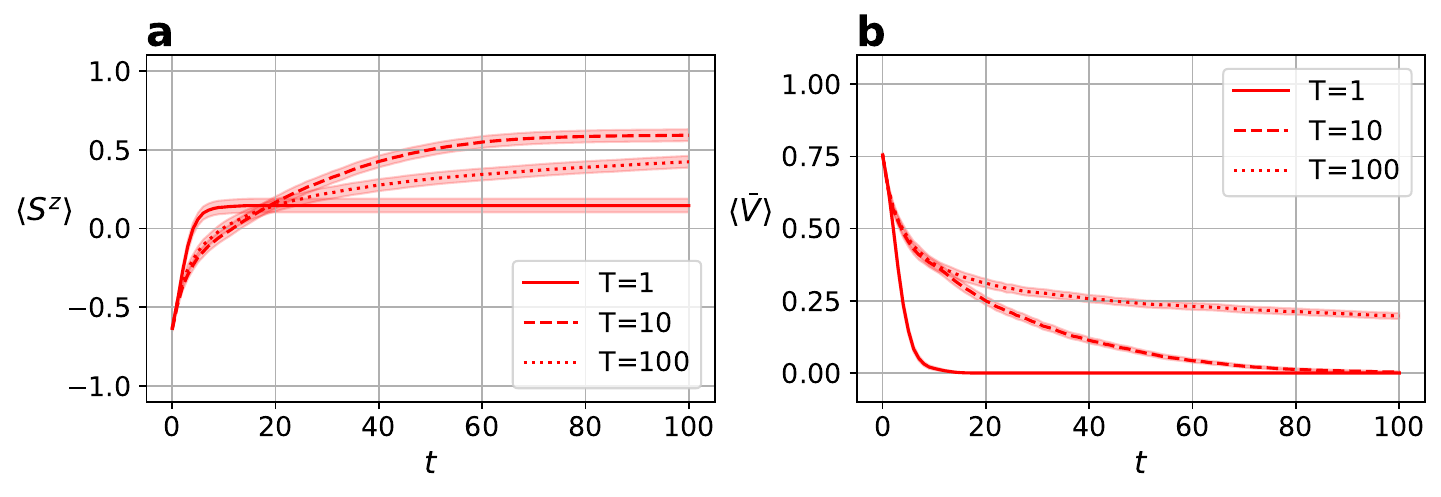}
        \caption{
        Time evolution of (\textbf{a}) the group performance $\langle S^z(t) \rangle$ and (\textbf{b}) the opinion diversity $\langle {\bar V}(t)\rangle $ for the SH initial condition at various values of the memory horizon $T=1$, $T=10$, and $T=100$. We have used the same parameter values $N=4$, $\alpha=\beta = 0.5$, $t_f = 100$  as for Fig.~\ref{fig:IC}. We find that as $T$ becomes larger, temporal changes slow down. For more details on the symbols and abbreviations, refer to the caption in Fig.~\ref{fig:IC}.    
        }
        \label{fig:T}
\end{figure}

\begin{figure}[t]
    \centering
        \includegraphics[width=0.9\textwidth]{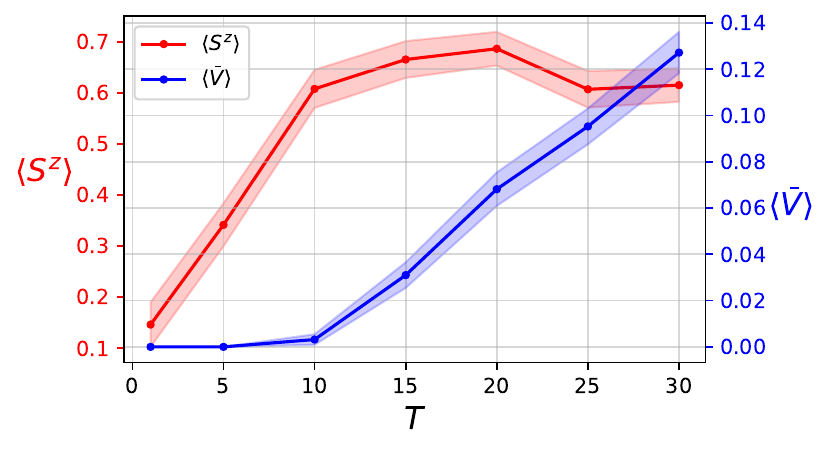}
        \caption{
       Plot of the group performance $\langle S^z(t) \rangle$ and opinion diversity $\langle {\bar V}(t)\rangle $ at $t = t_f = 100$ as a function of the memory horizon $T$ (see Fig.~\ref{fig:T}) for the initial condition SH. It is shown that there exists a maximum in the group performance at the intermediate value of $T$. For more details on the symbols and abbreviations, refer to the caption in Fig.~\ref{fig:IC}.
    }
        \label{fig:Tfinal}
\end{figure}

\begin{figure}[h]
    \centering
        \includegraphics[width=1\textwidth]{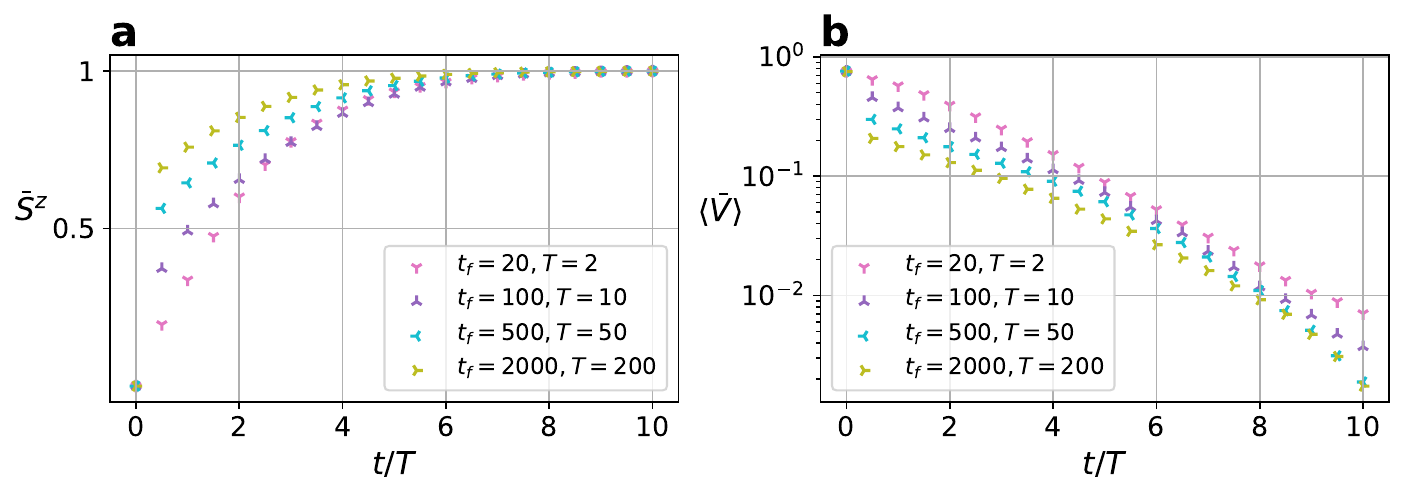}
        \caption{
        Scaling plots of (\textbf{a}) the normalized group performance defined as $\bar{S^z} \equiv \frac{\langle S^z \rangle-\langle S^z\rangle_\text{min}}{\langle S^{z}\rangle_\text{max}-\langle S^z\rangle_\text{min}}$ and (\textbf{b}) the opinion diversity $\langle {\bar V}(t)\rangle $, plotted against scaled time $t/T$ for the initial condition SH. Note that all data are measured at a fixed ratio of $t_f/T = 10$ and that symbol sizes do not represent standard errors. For more details on the symbols and abbreviations, refer to the caption in Fig.~\ref{fig:IC}.
        }
        \label{fig:scale}
\end{figure}

Next, we investigate how the memory horizon parameter $T$ affects the simulation results. Figure~\ref{fig:T} shows the time evolution of the group performance $\langle S^z(t) \rangle$ [Fig.~\ref{fig:T}(a)] and the opinion diversity $\langle {\bar V}(t)\rangle$ [Fig.~\ref{fig:T}(b)] up to the final time $t_f = 100$ for the SH initial condition at various values of the memory horizon $T=1$, $T=10$, and $T=100$ with $N=4$ and $\alpha=\beta = 0.5$. We focus on the initial condition SH, where all initial opinions are scattered only on the southern hemisphere ($S_i^z(0) < 0$). Although not shown here, we have verified that other initial conditions (NH and EH) display qualitatively similar behaviors to SH.

Again, it is found that the group performance increases and the opinion diversity decreases as time proceeds. However, the role played by the memory horizon $T$ is very interesting. When the memory horizon $T$ is very small like $T=1$, the group performance increases very fast in the early stage of group discussion, but it approaches a relatively worse group performance level in the later stage. On the other hand, when $T$ is sufficiently large ($T=100$), it appears that the temporal changes become much slower, as can be seen in Fig.~\ref{fig:T}(b) from a relatively large nonzero value of $\langle {\bar V}(t)\rangle$. This suggests that the value of $t_f = 100$ for $T=100$ is not sufficiently large for the system to arrive at equilibrium. However, we believe that an extremely large value of $t_f$ like $t_f = 1000$ is not realistic in a practical sense considering the time constraint in real school activities. Within this limitation of $t_f=100$, it is interesting to note that the group performance $\langle S^z(t=t_f) \rangle$ is the highest for intermediate value ($T=10$) of the memory horizon, as shown in Fig.~\ref{fig:T}(a). Figure~\ref{fig:Tfinal} summarizes our results for the $T$-dependence of the group performance and the opinion diversity at $t_f = 100$. It is to be noted that while the opinion diversity $\langle {\bar V} \rangle $ exhibits a monotonically increasing behavior with $T$, the group performance $\langle S^z \rangle$ does not; it shows a maximum around $T=20$. 

From our above observation of dynamic slowing down at large values of $T$, we explore the possibility of scaling the observed quantities by using the scaling variable $t/T$. As shown in Fig.~\ref{fig:Tfinal}, group members succeed in achieving a unified consensus answer at the final time $t_f=100$ for $T\lesssim10$. To examine this scaling behavior, we fix the ratio $t_f/T=10$ and plot the time evolution of the group performance and the opinion diversity versus $t/T$ as shown in Fig.~\ref{fig:scale}. This illustrates that both quantities appear to collapse at $t/T \gtrsim 8$ for large values of $T$. Such a finding suggests that as long as the ratio of $t_f/T=10$ is preserved, different combinations of $t_f$ and $T$ would yield equivalent outcomes if we focus on the equilibrium behavior. Consequently, our selection of $T=10$ and $t_f = 100$ with $t_f/T = 10$ can be used to simulate educational phenomena, considering the practical time constraints.

\subsection{Effect of group size}

\begin{figure}[t]
    \centering
    \includegraphics[width=1\textwidth]{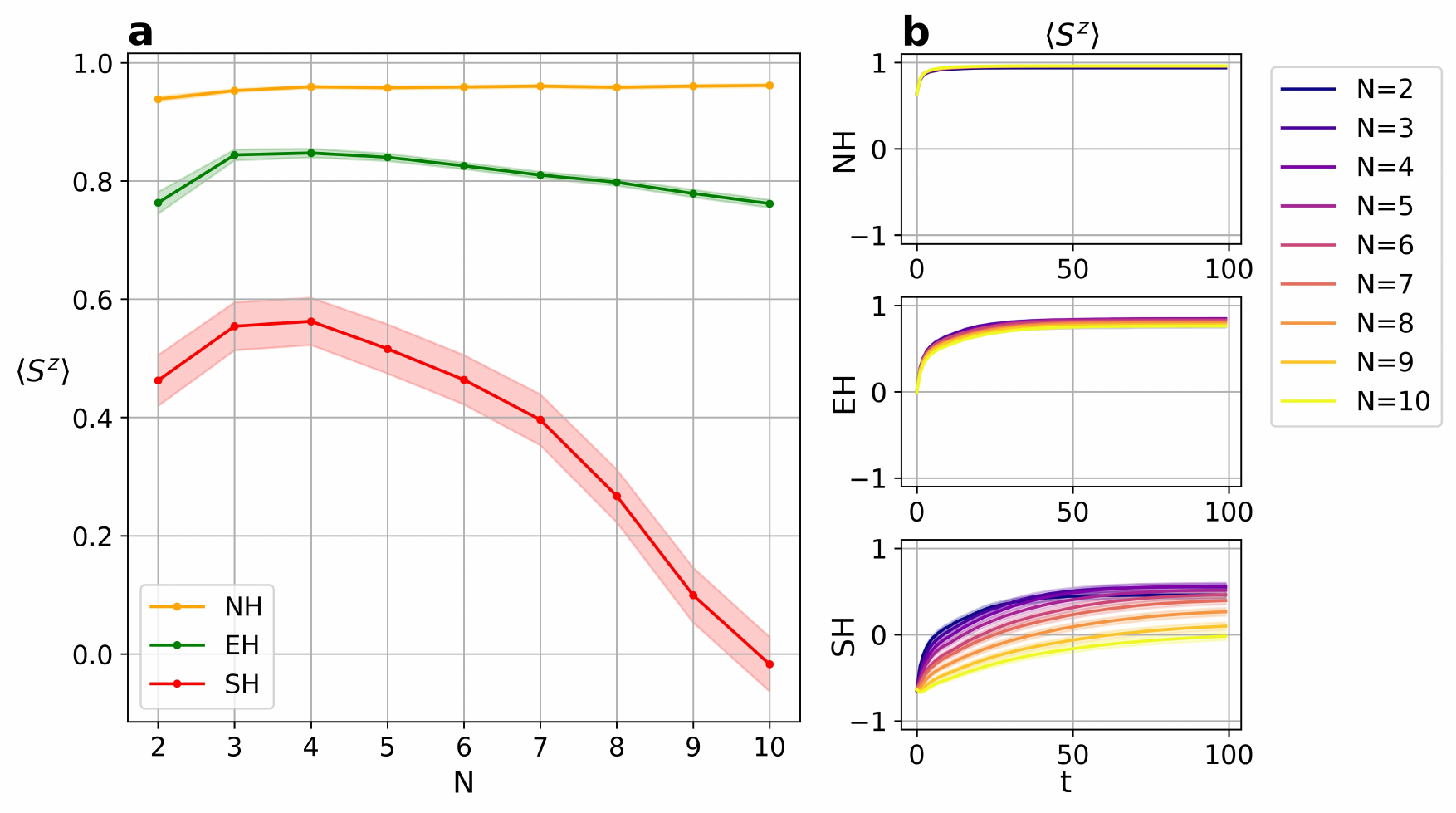}
    \caption{
        (\textbf{a}) Final group performance $\langle S^z(t=t_f) \rangle$ as a function of the group size $N$ for initial conditions NH, EH, and SH. (\textbf{b}) Time evolution of the group performance $\langle S^z(t) \rangle$ is displayed for each initial condition (NH, EH, and SH) with various system sizes. For more details on the symbols and abbreviations, refer to the caption in Fig.~\ref{fig:IC}.
        }
        \label{fig:N}
\end{figure}

From the above reasoning, we now fix $T=10$ and $t_f=100$. Although we find that $t_f=100$ is not sufficiently long enough to ensure the stationarity for larger system sizes like $N=10$, we choose to fix $t_f=100$ considering the practical time limitation in reality.

Figure~\ref{fig:N}(a) displays the final group performance $\langle S^z(t=t_f) \rangle$ versus the group size $N$ for initial conditions NH, EH, and SH. For NH, it is shown that the final group performance is not much affected by $N$, maintaining consistently high performance for all system sizes. In comparison, the outcome from the EH initial condition lies consistently lower than those of the NH initial condition, and it exhibits a gradually decreasing pattern with $N$ beyond $N=4$. Interestingly, for the SH initial condition, the final group performance displays a clear maximum around $N=3$ or $N=4$. Additionally, we find that the group performance for $N=2$ is worse than $N=3$, implying that too small group size hinders the group of students from finding the correct answer through group discussions. Our results indicate that group size does not significantly impact the group performance for NH and EH, but the choice of a proper group size can be a critical factor for lower-performing groups like SH. Overall, for all initial conditions tested in the present work, the optimal group size for better collaborative learning appears to be $N=3$ or $N=4$. In Fig.~\ref{fig:N}(b), we show the time evolution of the group performance for NH, EH, and SH through nine trajectories for different group sizes, respectively. It is shown that for the SH initial condition, the different random configuration of the initial opinions exhibits a largely different trajectory, which is reflected as a large error bar size in Fig.~\ref{fig:N}(a). The performance maximum around $N=3$ or $N=4$ is related to our choice of $\alpha = 0.5$ in Eq.~(\ref{eq:SG}) and we note that larger values of $\alpha$ shift the position of the performance maximum toward a larger group size $N$.

\subsection{Influence by initial conditions}

\begin{figure}[t]
    \centering
    \includegraphics[width=1\textwidth]{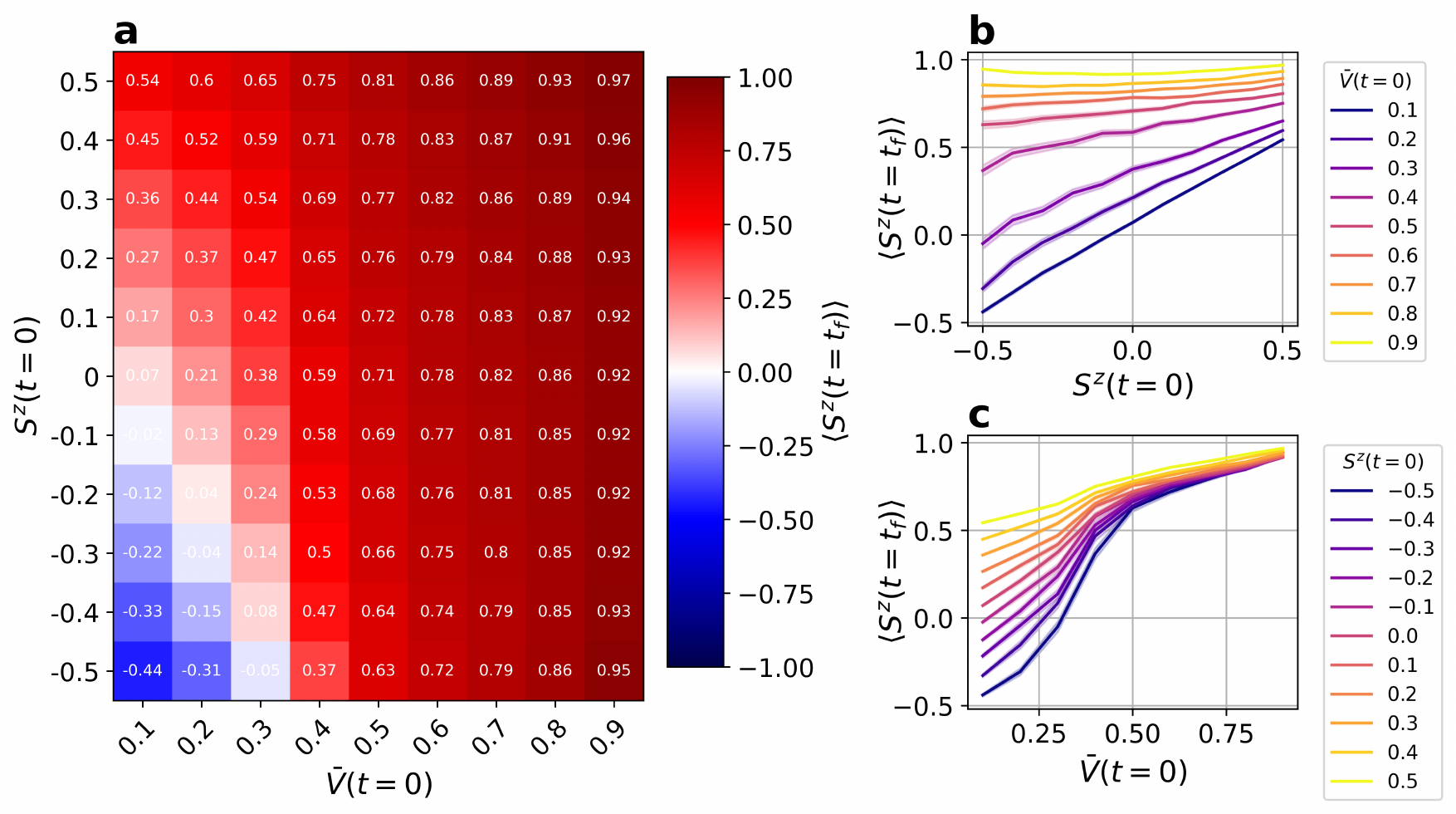}
    \caption{
        (\textbf{a}) Heat map for the final group performance $\langle S^z(t=t_f) \rangle$ in the plane of the initial group performance $S^z(t=0)$ and the initial opinion diversity ${\bar V}(t=0)$. (\textbf{b})  $\langle S^z(t=t_f) \rangle$ as a function of $S^z(t=0)$ at various values of ${\bar V}(t=0)$. (\textbf{c})  $\langle S^z(t=t_f) \rangle$ as a function of ${\bar V}(t=0)$ at various values of $S^z(t=0)$. For more details on the symbols, refer to the caption in Fig.~\ref{fig:IC}.
        }
        \label{fig:All}
\end{figure}

We next focus on the effects of different initial conditions on the final group performance in a more general context. We systematically generate random configurations of initial opinions as follows: We start from randomly scattered initial opinion vectors on the unit sphere, and make a random change of each vector until the target value [$\in (0.1, 0.2, \cdots, 0.9)$ with interval 0.1] of the initial opinion diversity ${\bar V}$ is achieved. We then rotate the whole set of opinions to achieve the target value [$\in (0.1, 0.2, \cdots, 0.9)$ with interval 0.1] of the initial group performance. Once we get the initial opinion vectors constrained to given values of $S^z (t=0)$ and ${\bar V}(t=0)$ for initial group performance and the opinion diversity, respectively, the system evolves in time as described in the Section~\ref{sec:2.3}. We produce 1,000 independent trajectories for each combination of $S^z (t=0)$ and ${\bar V}(t=0)$. Figure~\ref{fig:All}(a) shows the color-coded ensemble averaged final group performance $\langle S^z (t=t_f=100) \rangle$ and its values on the plane of $S^z (t=0)$ and ${\bar V}(t=0)$, obtained for $N=4$ and $T=10$.

For the top-right corner of Fig.~\ref{fig:All}(a), where both the initial group performance and initial group opinion diversity are sufficiently high, we expect the final group performance to be high as well.
On the other hand, the bottom-left corner represents the worst-case scenario, where both the initial group performance and opinion diversity are the worst, leading us to expect that the final group performance will also be the lowest.
Furthermore, we expect that for a given value of the initial group performance (the opinion diversity), the final group performance is an increasing function of the initial opinion diversity (the initial group performance) [see Figs.~\ref{fig:All}(b) and \ref{fig:All}(c)].

Our intuitive expectations are confirmed in Fig.~\ref{fig:All}, indicating the existence of a phase boundary between the low and high group performance phases. We report three principal findings: (1) For a fixed value of the initial opinion diversity, the final group performance at $t=t_f$ increases almost linearly with the initial group performance as clearly shown in Fig.~\ref{fig:All}(b). This suggests that the closer the initial opinions are to the correct answers, the higher the final performance of the group becomes. (2) For a fixed value of the initial group performance, the final group performance is again an increasing function of the initial opinion diversity, as shown in Fig.~\ref{fig:All}(c). (3) Another extremely interesting observation one can make in Fig.~\ref{fig:All}(c) is that the final group performance does not depend strongly on the initial group performance when the initial opinion diversity is sufficiently high ($\bar{V}(t=0) \gtrsim 0.5$). This suggests that the threshold $\bar{V}(t=0) \approx 0.5$ represents the optimal level of opinion diversity which is necessary to achieve the best outcomes in group discussions, regardless of group composition. We thus conclude that forming groups with diverse opinions to start with is crucial for the group to approach the correct answer.

We found that the phase change observed in Fig.~\ref{fig:All} arises from the non-linear weight applied to the mutual influence term in the RHS of $\vec{G_i}(t)$ in Eq.~(\ref{eq:SG}). As we have explained, students tend to be more influenced by opinions similar to theirs, and the presence of initial opinion diversity can improve group performance through peer discussions. Therefore, when opinions are diverse, they tend to move toward the $+z$ direction and eventually converge once they become similar. However, if the initial opinion diversity is low, the scaffolding effect is too weak to achieve higher scores. This explains why the sudden state transition occurs around $\bar{V}(t=0) \approx 0.5$.
We further verified that if this term is instead formulated as $\sum_j \vec{S_j}$, the final group performance increases only linearly and no collapse occurs even with increased initial opinion diversity, contrasting with the results in Fig.~\ref{fig:All}(c). This is because, in this case, opinions are updated to the average opinion vector but are biased toward the $+z$ direction due to the scaffolding effect.

\section{Summary and Discussion}
In this paper, we have proposed a model that simulates the educational phenomenon in which group discussions lead to the correct answer. In our model, a student's opinion is represented as a three-dimensional unit vector, which evolves through influence from peers with the memory effect. We have also modeled the scaffolding effect that guides group members toward the correct answer in analogy to an external field in statistical mechanics models. While several previous studies have examined the social learning process within a classroom~\cite{pritchard2008mathematical, nitta2010mathematical, castellano2009statistical, bordogna2001theoretical, bordogna2002cellular, bordogna2003simulation, burini2016collective, ormazabal2021agent}, they have not clearly demonstrated significant performance improvement through peer-collaborative activities that exclude teacher assistance and informational resources, especially when initial opinions largely deviate from the correct answer and vary among students. We have shown from our model study that the opinion diversity $\langle {\bar V}(t)\rangle$ defined in Eq.~(\ref{eq:V}) within a group leads to an improvement in the group performance $\langle S^z (t) \rangle$. 

Our numerical investigations have led us to the following findings. The group performance $\langle S^z (t) \rangle$ increases through discussions—--even when none of the group members initially knows the correct answer—--as long as their initial opinions sufficiently differ. It has also been observed that the enhancement of the group performance is larger when the initial opinions are more diverse, highlighting significant advancements in the performance of heterogeneous groups.
Moreover, we have found that the improvement in group performance is more influenced by initial opinion diversity $\langle {\bar V}(t=0)\rangle$ than by initial group performance $\langle S^z (t=0) \rangle$: When $\langle {\bar V}(t=0)\rangle \gtrsim 0.5$, the final group performance $\langle S^z (t=t_f) \rangle$ tends to converge at a higher level regardless of $\langle S^z (t=0) \rangle$.
But even under this condition, a higher initial group performance still leads to a slightly higher final group performance, which is related to the findings in Section~\ref{sec:3.1}.
We have confirmed that these findings are consistent with two key results from a previous study on peer-collaborated activities~\cite{lee2022effect}: (1) Even when two groups start with similar initial group scores, their performances after discussion differ; groups with more diverse initial answers exhibit greater improvement. This suggests that forming groups with diverse opinions is crucial for enhancing group performance. (2) A higher initial group performance contributes to better final performance, which is a rather straightforward result.

The memory horizon $T$ has been shown to affect both temporal dynamics and performances: The shorter memory horizon makes the consensus occur earlier but the final group performance becomes worse. These results align with the theoretical study showing that the memory horizon of individuals is inversely proportional to the speed of final opinion formation~\cite{liu2023memory} and with an empirical study showing that groups with larger memory capacities perform better in collaborative learning~\cite{du2022impact}. As the memory horizon $T$ is increased, the increase of the group performance has been found to become slower but the final performance is enhanced. This result agrees with March's model~\cite{march1991exploration}, which suggests that if adaptive processes (e.g., modifying opinions during discussions) occur too rapidly, it might be effective for learning in the short term but not in the long term.

We have also reported that the group performance $\langle S^z (t) \rangle$ strongly depends on the group size, especially when the initial group performance is low. For all initial conditions we have tested in our work, there appear to exist the optimal group sizes for fostering effective collaborative learning. Early studies on collaborative learning in physics have also found that small group sizes of three to four are effective~\cite{heller1992teaching1, leinonen2017peer}. In contrast, we have found that too small group size like $N=2$ fails to suitably manage conflicts within a group and leads to worse group performance, aligning with previous research~\cite{leinonen2017peer, alexopoulou1996small}. Moreover, it has been shown that the performance of large groups tends to be overestimated, with negative side effects such as lower satisfaction and reduced involvement in general educational situations~\cite{hill1982group}.

In this study, we mathematically defined the scaffolding effect with group size and opinion diversity.
Although there are challenges in rigorously validating our model with real data, our findings remain consistent with previous educational observations~\cite{lee2022effect,singh2005impact,smith2009peer,brundage2023peer}. One limitation is the lack of real temporal data, which makes it difficult to directly estimate the parameters $\alpha$ and $\beta$. However, this does not diminish the model’s ability to capture key dynamics of group discussions. Another challenge is the difference in performance measurement: The previous study~\cite{lee2022effect} assessed group performance measured by the probability that the group finds the correct answer, whereas our study uses the average partial scores of group members. This difference is due to how we represent student's opinion and performance.

To enhance model validation, introducing large language model (LLM) agents presents a promising approach. Recent studies suggest that LLM agents can simulate real humans when provided with appropriate contextual information~\cite{argyle2023out, aher2023using}. With targeted prompt engineering, these agents could function as substitutes for real students, enabling more systematic validation of our model. Future research could explore this avenue to bridge the gap between theoretical modeling and empirical validation.

Furthermore, while empirical and theoretical findings confirm the presence of scaffolding effect by peers in collaborative learning~\cite{bordogna2001theoretical,lee2022effect,singh2005impact,smith2009peer,brundage2023peer,heller1992teaching1, leinonen2017peer, alexopoulou1996small}, the process itself has not been thoroughly studied using a mathematical approach. Therefore, as a future work, it would be valuable to study the scaffolding effect among peers mathematically, for instance, through the framework of the game theory.

Lastly, although our model only addresses ‘discussion’, we expect that it could also be adapted to analyze more complex scenarios, such as ‘debate’ or ‘dispute’. Binary or multi-state voter models have been used to study disputes with the assumption that all options are equivalents~\cite{starnini2012ordering, nowak2021discontinuous}. For example, in scenarios with opinions A, B, and C, these models often disregard the fact that some opinions (e.g., A and B) might be closer (more similar) than others (e.g., A and C). Therefore, our model's approach, which represents opinions of agents on a unit sphere, could offer a precise way to position each opinion similarly to reality.

\section*{Acknowledgements}
This work was supported by the National Research Foundation (NRF) of Korea through Grant Numbers RS-2023-00214071 (J.S.) and RS-2024-00341317 (B.J.K.).

\section*{Data availability}
The data and code supporting the findings of this study are openly accessible through the following repository: \href{https://github.com/jabamseo/24col_study}{https://github.com/jabamseo/-24col$\_$study}.

\end{document}